\documentclass[a4,12pt,elsart]{article}
\usepackage{graphics}
\usepackage{epsfig}
\def\BC{bounded confidence }
\def\sfn{scale free networks }

\begin{document}
\title{Persuasion dynamics}

\author{
 G{\'e}rard~Weisbuch$^{*}$ \\
Guillaume Deffuant$^{**}$ and Fr{\'e}d{\'e}ric Amblard$^{**}$ \\
$^{*}$Laboratoire de Physique Statistique\footnote{
Laboratoire associ{\'e} au CNRS (URA 1306), {\`a} l'ENS et
 aux Universit{\'e}s Paris 6 et Paris 7}
\\
     de l'Ecole Normale Sup{\'e}rieure, \\
    24 rue Lhomond, F-75231 Paris Cedex 5, France. \\
$^{**}$Laboratoire d'Ing{\'e}nierie pour les Syst{\`e}mes Complexes (LISC)\\
Cemagref - Grpt de Clermont-Ferrand\\
24 Av. des Landais - BP50085\\
F-63172 Aubi{\`e}re Cedex (France)\\
  {\small     {\em email}:weisbuch@lps.ens.fr}\\
 }
\maketitle

\abstract{We here discuss a model of continuous
 opinion dynamics in which
agents adjust continuous opinions as a result of
random binary encounters whenever their  difference in opinion
 is below a given threshold. We concentrate
on the version of the model in the presence of few extremists
which might drive the dynamics to generalised  extremism.
The intricate regime diagram
is explained by a combination of meso-scale features
involving the first interaction steps}

%%\titlerunning{Adjustment and social choice}
%%\author{G{\'e}rard~Weisbuch\inst{1}
%%\and Dietrich Stauffer\inst{2}}
%
%%\authorrunning{G{\'e}rard~Weisbuch et al.}

%\institute{Laboratoire de Physique Statistique\footnote{Laboratoire
% associ{\'e} au CNRS (URA 1306), {\`a} l'ENS 
%et aux Universit{\'e}s Paris 6/7, e-mail: weisbuch@lps.ens.fr }
%de l'Ecole Normale Sup{\'e}rieure,
%24 rue Lhomond, F 75231 Paris Cedex 5, France

%%\maketitle              % typesets the title of the contribution

%\begin{abstract}
%\end{abstract}

\section{Introduction}
\label{sec:intro}

   The present paper is a follow-up in a series of publications on
continuous opinion dynamics and "extremism" \cite{dext}.

  Social psychology is often concerned with the outcome of
collective decision processes in connection with individual
cognitive processes and the actual dynamics of 
opinion exchanges in meetings.
The issue of whether extremist or moderate opinions are adopted 
in commissions is thoroughly discussed in Moscovici and Doise \cite{mosco}.

A connection with statistical physics and the Ising model
was soon established, for instance by Galam \cite{Galam}
who collaborated with
Moscovici. They considered binary opinions as in 
the vast majority of the literature on binary social choice
\cite{Follmer}. 

  The approach here is different and rests on the fact 
that certain choices imply continuous opinions; typical examples
are the evaluation of economic profits among different possible choices
\cite{images}, or how
to share profits after a collective enterprise (hunt, agriculture
etc.)\cite{py}.
In the early literature on committees, 
opinions were simply supposed to influence each other 
in proportion to their difference. The described dynamics was 
equivalent to heat diffusion, and resulted in uniformisation
around some average opinion.

The notion of an interaction threshold, based on experimental 
social psychology, was proposed by Chattoe and Gilbert \cite{chat},
and introduced in models by Deffuant etal \cite{Deffuant}
and Hegselmann and Krause \cite{Hegselmann}.
Two individuals with different opinions only influence
each other when their difference in opinion
is lower than a threshold. The outcome of the dynamics
 can then be clustering rather to uniformity.
(A series of models of cultural diffusion
first introduced by Axelrod and followers \cite{axel}
belong to the same class: cultures are represented by
vectors of integers
which are brought closer by interactions under certain conditions
of similarity; these authors studied how these conditions
influence the outcome of the dynamics, uniformity
versus diversity. Integer variables facilitate analytical
approaches \cite{axel},\cite{red} via master equations). 

   Fascinating results were obtained in the "extremism"
model of Deffuant etal \cite{dext}:
 when interaction thresholds are unevenly distributed,
 and in particular when agents with extreme opinions
are supposed to have a very low threshold for interaction,
extremism can prevail, even when the initially extremist agents
are in very small proportion. The so-called "extremist model"
can be applied to political extremism, and a lot of the heat 
of the discussion generated by these models relates
to our everyday concerns about extremism. But we can think of 
many other situations where some "inflexible" agents are more sure 
about their own opinion than others. Inflexibility can arise for instance:

\begin{itemize}
\item Because of knowledge; some agents might know the answer
while others only have opinions; think of scientific knowledge and the 
diffusion of new theories;
\item Some agents might have vested interests different
from others.
\end{itemize}
  Although the model has some potential for many applications,
we will here use the original vocabulary of extremism.

  Several subsequent papers \cite{ambdef}
checked the genericity of these results
for different interaction topologies (well-mixed systems
versus social networks represented by many variants between 
lattices and random nets) 
and for different variants of the distribution of interaction intensities
(see further equation 3).
Clustering and the possibility of extremist attractors
were shown to be generic, but the phase diagrams between
different dynamical regimes can be rather intricate with co-existence
regions depending on parameter values.

The purpose of the present paper is to increase our understanding 
of these phase diagrams using simpler conditions for simulation.
(Unfortunately, were are still not very advanced
in formal derivations).
 The subsequent section describes
the models, the simulation techniques and the monitoring 
of the results.  We first deal with models where only
one "extremist" is present. Full mixing and lattice topologies
are then studied. More intricate situations with 
many extremists can be understood from the one-extremist case.
 
\section{Models and simulations}

\subsection{The basic model}
The most basic model later called bounded confidence model
was introduced by Deffuant etal \cite{Deffuant}. It supposes  an
initial distribution of agents with scalar opinion $x$.

At each time step any two
randomly chosen agents
meet: they re-adjust their opinion when their difference in
 opinion is smaller in magnitude than a threshold $d$.
Suppose that the two agents have opinion $x$ and $x'$.
$Iff$ $|x-x'|<d$ opinions are adjusted according to:
\begin{eqnarray}
  x &=& x + \mu \cdot (x'-x) \\
  x' &=& x' + \mu \cdot (x-x')
\label{eq1}
\end{eqnarray}
\begin{figure}[htbp]
\centerline{\epsfxsize=60mm\epsfbox{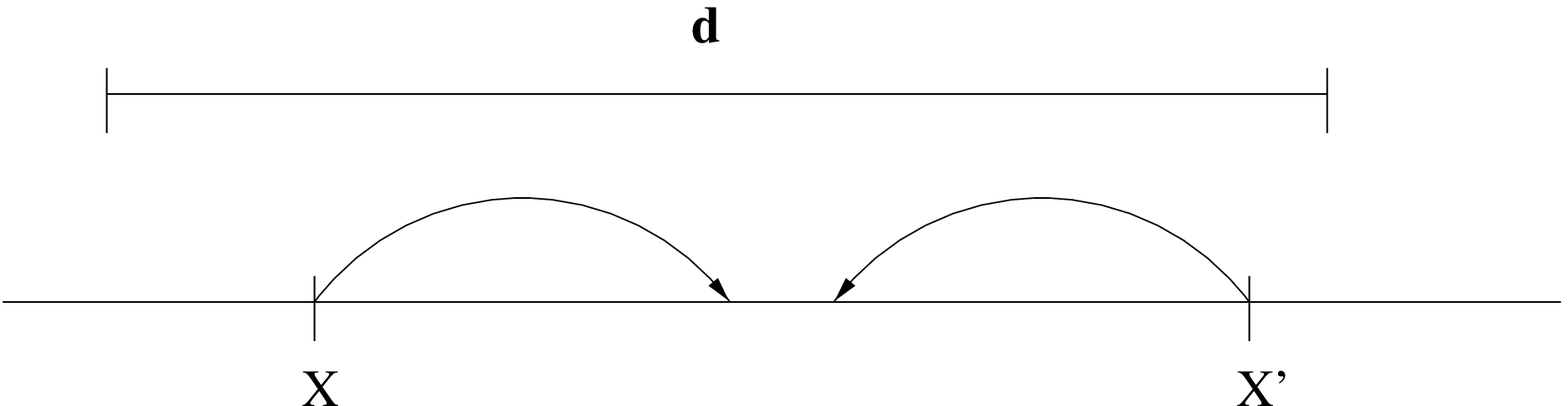}}
\end{figure}

 where $\mu$ is the convergence parameter
whose values may range from 0 to 0.5.
The rationale for the threshold condition is that agents only
interact when their opinions are already close enough;
otherwise they do not even bother to discuss.

The basic model uses 
\begin{itemize}
\item 
a threshold $d$ 
 constant in time and across the whole population;
\item  a complete mixing hypothesis;
\item  and 
 a random serial iteration mode.
\end{itemize}

The threshold can be interpreted as "openness",
tolerance or as some uncertainty in opinion.

 The choice of the random serial iteration mode
models opinion diffusion in a large population
which agents encounters each other in small groups
such as pairs. By contrast Hegselmann and Krause \cite{Hegselmann}
 chose parallel iteration since their approach derives from 
earlier literature modeling formal meetings.

 Computer simulations show that the distribution of opinions
evolves at large times towards clusters of homogeneous opinions
(both iteration modes yield similar clusters under the 
used conditions).

 For large threshold values ($d>0.3$) only one single cluster
is observed at the average initial opinion (consensus).
For lower threshold values, several large clusters are observed.
 Consensus is then NOT achieved when thresholds are low
enough. The number of clusters varies as the integer part of $1/2d$
\cite{Deffuant},
to be further referred too as the $1/2d$ rule.

Some recent literature by Stauffer and collaborators
\cite{HMO} consider as clusters any group of opinions
however small (even of size one). Counting all these groups
yield higher figures scaling with $N$ the number of agents. 
We here only monitor large clusters which size
is a finite fraction of the number $N$ of agents.
We don't care about the existence of isolated ``outliers''.
(because of the randomness of the iteration process some
agents are selected at later times and remains
 as ``outliers'' outside the 
main clusters). The ``generic'' results we here
refer to, such as the $1/2d$ rule, apply to the large clusters.

  Rewriting the opinion updating equation as:
\begin{eqnarray}
  x &=& x + \mu \cdot f(x'-x) \cdot (x'-x),
\end{eqnarray}
the \BC model supposes a square amplitude of the 
interaction function $f(x'-x)$ when $|x-x'|<d$.
 Smoother shapes 
(such as trapezoidal\cite{dext} or bell-shaped \cite{Val}) were also
proposed for $f(x'-x)$. The simulations show that
 the main dynamical features are conserved 
with these smoother interaction functions.

\subsection{Extremism}
 The model for extremism introduce by Deffuant etal \cite{dext}
is based on two more assumptions.

\begin{itemize}
\item 
A few extremists with extreme opinions at 
the ends of the opinion spectrum and with very low threshold
 for interaction are introduced.
\item 
 Whenever the threshold allows interaction, both opinions
and threshold are readjusted according to similar expressions.
\end{itemize}

  $Iff$ $|x-x'|<d$
\begin{eqnarray}
  x &=& x + \mu \cdot (x'-x) \\
  d &=& d + \mu \cdot (d'-d)
\end{eqnarray}
A symmetrical condition and equations apply to the other agent
of the pair with opinion $x'$ and tolerance $d'$ 
but when thresholds are different the influence 
can be asymmetric: the more "tolerant" agent (with larger $d$)
can be influenced by the less tolerant (with smaller $d$)
while the less tolerant agent is not. This ``effective''
asymmetry is responsible for the outcome of ``extremist''
attractors.

\subsection{Simulation methods and displays}
  Computer simulations are run
according to standard conditions.
\begin{itemize}
\item Initial conditions:
  uniform distribution of opinions among $[0,1]$
among $N$ agents with initial threshold $dl$.
Among these a few agents are extremists, with opinions
at the extreme of the opinion spectrum and with 
initial threshold $de<dl$.   
\item At each time step, one randomly selected pair
is chosen and agents are updated according to equations 4-5
whenever the condition on threshold is fulfilled.
\item Simulations are run until an approximate 
state of equilibrium is reached: we here consider that equilibrium is 
reached when opinion and tolerance histograms of 101 bins are stable. 
\end{itemize}
  The main parameters are the number and initial tolerance of 
extremists, and the initial tolerance $dl$ of the other agents.
  Variants include different interaction networks,
and different interaction functions $f(x'-x)$. 

 We usually first check opinion and tolerance dynamics by time plots
of single simulations \cite{dext}.
 These time plots are clouds of points representing
at each time step the opinions and tolerance of those agents chosen for 
eventual updating versus time along the x axis.

\begin{figure}[h]
\centerline{\def \epsfsize#1#2{0.5#1}\epsfbox{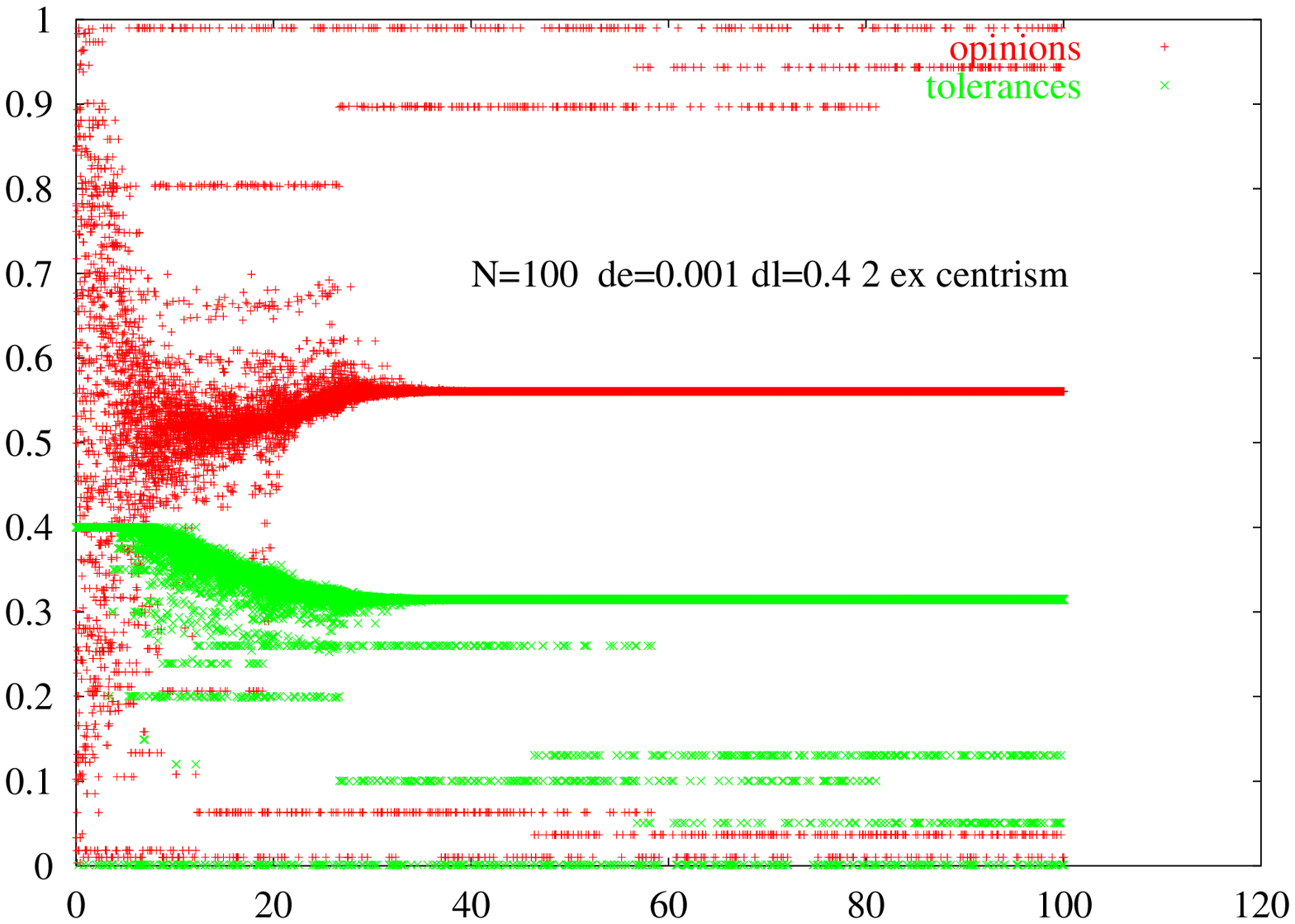}
\hskip 0.3 cm\epsfbox{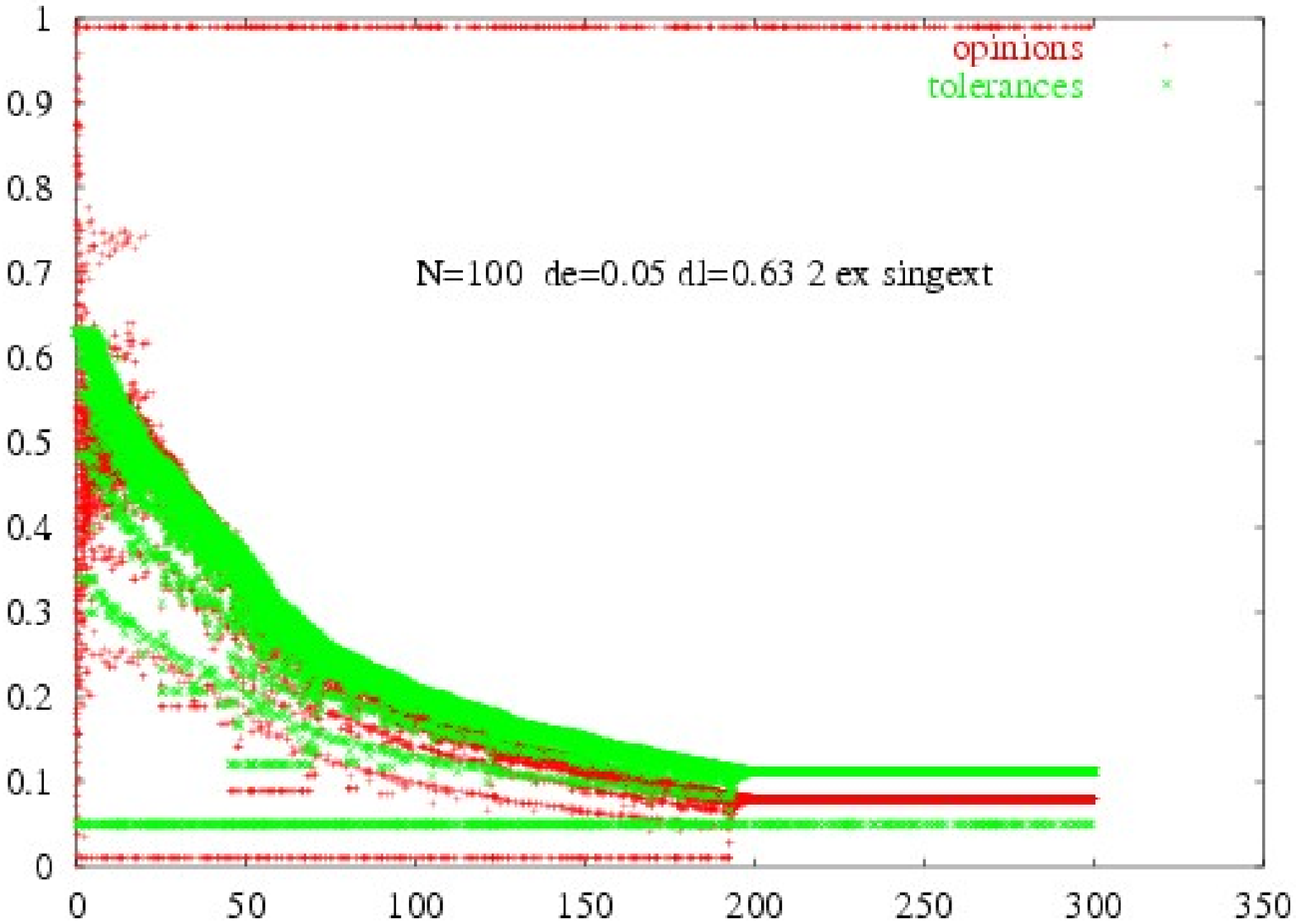}}
\caption{Time plots of opinion (red '+') and tolerance (green 'x')
dynamics exhibiting a "centrism" attractor on the left frame
and a single extremist cluster on the right frame.
Time is given in the average number of updating per agent.
The number of agents is $N=100$, extremists'
tolerance is $de=0.001$ and two opposite extremists are initially present.
Any pair of agents can a priori interact.
The centrism attractor is obtained when the initial centrist
tolerance is $dl=0.4$ and the extremist attractor when $dl=0.63$  
 }
\end{figure}

  The time plots display different dynamical regimes
according to the eventual predominance of the extremists:
sometime they remain isolated and most agents cluster 
as if there were no extremist (e.g as represented on figure 1
left frame); otherwise extremism
prevails and most agents cluster in the neighborhood
of one (e.g as represented on figure 1
right frame) or both extreme. Convergence characteristic time
differs: convergence is fast for the centrism attractor and 
slow for the extremism attractor. The ratio in convergence time
is approximately the ration in the initial fraction of 
centrists and extremists.  

  Which attractor is reached depends mainly upon the 
parameters of the simulation (number and initial tolerance of 
extremists, and the initial tolerance $dl$ of the other agents).
A simplified conclusion is that some kind of extremism 
prevails for larger values of the tolerance
of initially non-extremist agents when $dl>0.5$,
and centrism when $dl<0.5$. In other words, the outcome of the
dynamics is largely determined by the tolerance
of the non-extremists agents.
Systematic studies show the existence of parameter
regions where several attractors can be reached depending upon
the specific initial distribution of opinions and upon the specific choice 
of updated pairs.

  Deffuant etal \cite{dext} papers are filled with two dimensional 
regime diagrams coded according to a variant of the 
Derrida-Flyvbjerg parameter \cite{Derrida} defined as:

\begin{equation}
  Y=\sum\limits_{i}{\frac{n_i}{N}}^2
\end{equation}

This sum of the square of the fraction of number $n_i$ of agents
in each cluster $i$ roughly represents the inverse of a weighted number of
clusters. Particular choices of monitored Derrida-Flyvbjerg parameters
allow to separate dynamical regimes of attraction towards center,
clustering and attraction towards one or both extremes. 

  But these diagrams although comprehensive in terms of parameter ranges and
  averaging over many initial conditions are difficult to interprete and we
  use here a more direct approach. We only vary one parameter
along the x axis, most often the initial large tolerance $dl$.
The y axis code the histogram of attractor clusters by vertical
bars. The magnitude of the bar represents
how many agents are in the asymptotic cluster(s).
Clusters made of one agent are most often discarded to make diagrams 
more readable.
 The position of the bar represents either the opinion
 or tolerance of agents in that
  cluster. Each bar only gives the result of one simulation.

Coexistence regions appear as $dl$ intervals on which large fluctuations
are observed in the cluster positions. Probabilities of either
regime are evaluated from their frequency of observation 
on any intervall. By contrast, pure regimes 
yield regular variations of cluster positions. 

We also show the asymptotic patterns of opinions and tolerance
for lattice topologies with color coding.

\section{Single extremist regimes}

\subsection{Single extremist with full mixing topology}

  To easily gain some insight, let us start
from a rather extreme case: one single extremist agent chosen with
initial opinion 0.99 and 0.001 tolerance. The topology is full mixing.
Large simulation times are used (10 000 iterations per agent)
 to ensure convergence under every simulation condition (figure 2). 

\begin{figure}[ht]
\centerline{
\epsfxsize=80mm
\epsfig{file=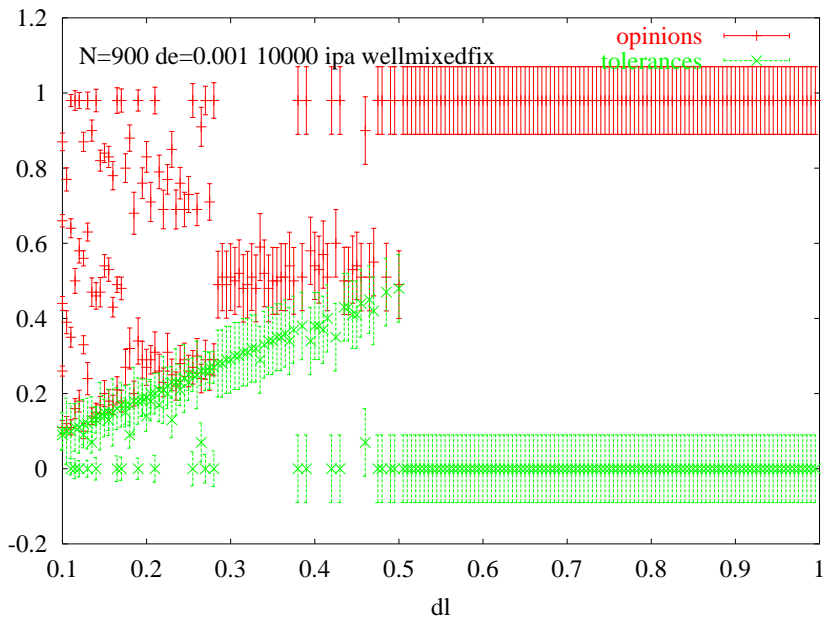,width=120mm}}
\caption{Histograms of asymptotic clusters.
 The y axis code the histogram of attractor clusters by vertical
bars which magnitude represents
how many agents are in the asymptotic cluster. The
  position of the bar represents either the opinion (red) or the 
tolerance (green) of the agents in that
  cluster. The horizontal axis gives the initial tolerance parameter of the
"centrist'' agents. One single extremist present,
 $N=900$, $de=0.001$, average number of iterations per agent
10 000, any pair of agents can a priori interact}
\end{figure}

 The regime diagram (figure 2) clearly shows that the centrist agents 
are all attracted by the extremist
when their initial tolerance $dl$ is above 0.5:
they gather in a cluster of opinion 0.99 and tolerance 0.001.
The interpretation is straightforward:
For this low value of extremist tolerance, interaction between
the extremist and centrists are asymmetric (as
we checked by measuring asymmetric interactions during
the simulation).
The extremist acts as a fixed source
of extremism, formally equivalent to a heat source at constant temperature
(equations 1 and 2 can be thought as a randomly discretized version
of a Euler relaxation algorithm solving a diffusion equation \cite{numrec}).
Opinion is here the equivalent of temperature.

Below $dl=0.5$ the influence of the extremist decreases
and agents cluster near the center opinion keeping roughly their initial
tolerance. When $dl<0.27$, the diagram show the same increase in cluster 
number that can be observed in the absence of extremist (the "1/2d rule"),
except for a partial extremism clustering below $dl=0.27$
 which is easily understood. 

The region $0.37<dl<0.52$ is a co-existence region where both regimes,
centrism or extremism can be observed, depending upon initial conditions
and pair sampling.

\subsection{Single extremist with square lattice topology}

\begin{figure}[ht]
\centerline{
\epsfxsize=80mm
\epsfig{file=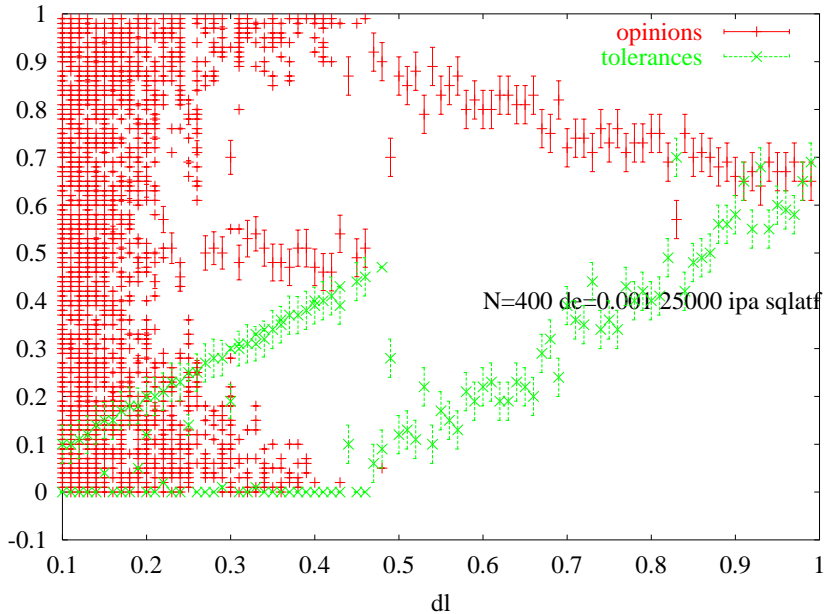,width=120mm}}
\caption{Histograms of asymptotic clusters for 
interactions across a square lattice topology.
Same coding as the previous figure. One single extremist present,
$N=400$, $de=0.001$, average number of iterations per agent
100 000. Single element clusters are exceptionnally represented
 on this figure, in order to check that the initial extremist himself
moves towards centrism at large $dl$ values. }
\end{figure}

  In many cases we expect interactions to occur
across some social network. Such would be the case
for political discussions, especially in the absence
of an open discussion forum. Many model topologies
of social networks have recently been proposed.
 We here report simulation results
for square lattices, when interactions are 
only possible among nearest neighbours (each
node can only interact with his 4 neighbours).
The boundaries of the lattice are connected
to each other: the diagrams represent in fact
the unfolding of a torus. 

 Although the regularity of connections
on a square lattice make it a poor candidate to model
a social network, the existence of short interaction
loops is shared with many empirical social nets. 
But again, the purpose of this paper is to increase our
understanding of the dynamics of more complicated cases
and the possibility to observe patterns determined our
choice of lattice topology. The relevance of the results
to other topologies will be further discussed. 

  The main difference in dynamics between well mixed systems and
the square lattice structure appears in the $dl>0.5$ region.
For values of $dl$ just above 0.5, the extremism regime seems to 
re-appear (see figure 3).
A closer examination of the dynamics, (see figure 4)
shows attraction towards extremism proceeds locally
on the lattice in the neighborhood of the extremist.
This spatial diffusion process is not the same as
the emergence of single sided extremism in well-mixed 
systems as described in Deffuant etal \cite{dext}:
in well-mixed systems one first observe a convergence towards
a attractor with centrist opinion and low tolerance
 which is often unstable and slowly evolves towards 
extremism.

\begin{figure}[ht]
\centerline{\def \epsfsize#1#2{0.8#1}\epsfbox{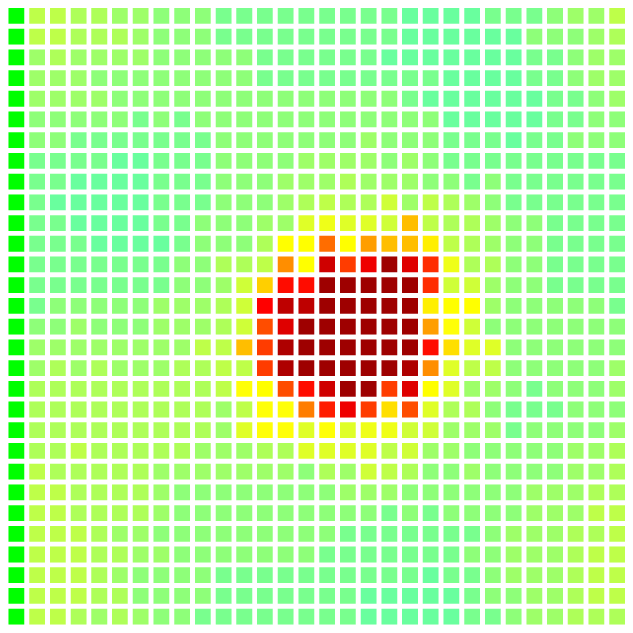}
\hskip 0.3 cm\epsfbox{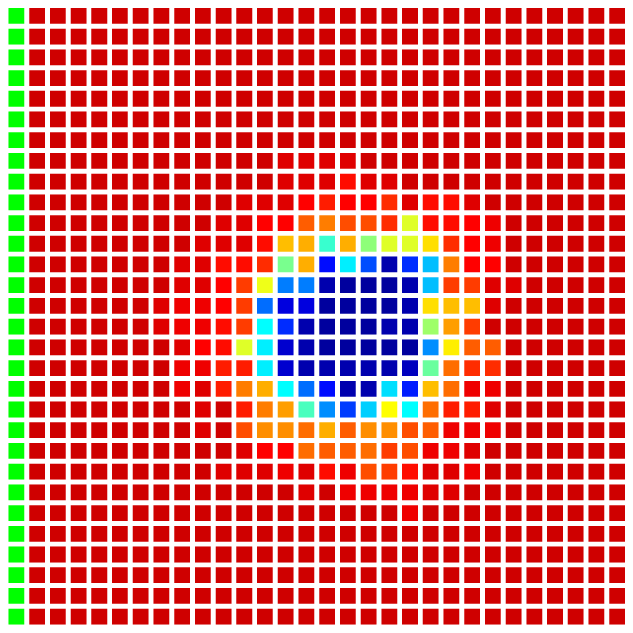}}
\caption{Propagation of extremism from a single initial extremist
at the center of the lattice.
Left frame: Opinions after 27 iterations per site 
. Color scale: deep blue 0,
light green 0.5, brown 1.
 Right frame: Tolerances after 27 iterations per site 
. Color scale: deep blue 0,
brown $d_l=0.55$, initial centrist's tolerance.
Initial extremist tolerance 0.001.}
\end{figure}

 But for large value of $dl$, 
the influence of extremist seems to weaken. 
Let us see why.
The propagation of extremism proceeds from
the initial extremist is initially asymmetric,
since the initial extremist and the new ``converts''
are separated from their
neighbours down stream by an opinion gap
intermediate between the two tolerances.
But as diffusion proceeds, the opinion difference  
decreases below the low tolerance of extremists
thus allowing symmetric interactions to occur,
as we checked during the simulations.
Initial and converted extremists
 become fully coupled by symmetric diffusion dynamics and their opinion
is also influenced by those of their centrist neighbours.
 The position of the final
cluster reflects this balance of influence. Initially
more tolerant agents (large values of $dl$)
are faster attracted towards the extremist, and their increased 
number favours the cluster evolution toward the center in opinion and towards
$dl$ in tolerance.(By contrast, in the case of full mixing,
all agents can be attracted by the extremist when it acts
as a source; there are no screening shells
in the vicinity of the extremist). 

 The effect is density dependant: for a smaller lattice,
$N=100$, the deviation of extremism towards the center is
much weaker (the opinion cluster is at 0.9 rather than 0.7 at
$dl$=0.99), for the same values of the initial tolerances.
  In other words, this effect is not a standard seed effect
as usually observed in phase transitions where a single seed
is able to drive all the system to another phase. 

  The same coexistence region when $0.37<dl<0.52$ with the 
occurence of either type of attractor is observed 
with the lattice topology,
as when all interactions are a priori possible.

 \section{Simulation with several extremists}

\subsection{Low extremist density and full mixing}

 Deffuant etal \cite{dext}  report the existence of several
dynamical regimes for the full mixing case
in the presence of extremists of both kind:
\begin{itemize}
\item When $dl<0.5$, extremists are not important and clustering follow
the standard $1/2d$ rule. 
\item  When $dl>0.5$ they determine the dynamics:
\begin{itemize} \item at high extremist initial density,
 clusters of extremists appear at both end of the spectrum; 
\item at low extremist density, instabilities often arise,
and the system might evolve in a single asymmetric extreme
attractor at one end of the spectrum, or even reach an attractor 
with centrist opinion but extremist low tolerance.
\end{itemize}
\end{itemize}

All time plots and regime diagrams are given in their
paper \cite{dext}.

\subsection{Low extremist density on a square lattice topology}

 \begin{figure}[ht]
\centerline{
\epsfxsize=80mm
\epsfig{file=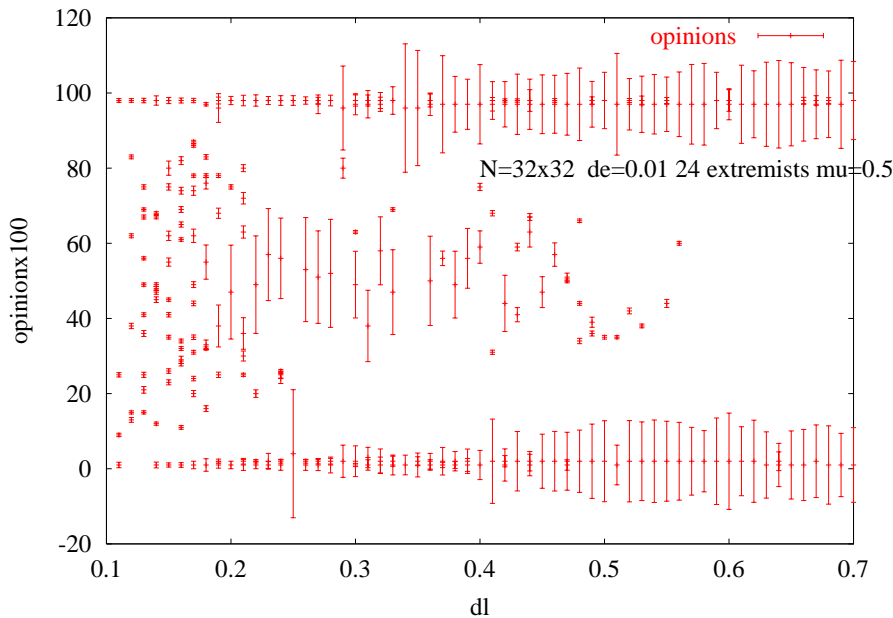,width=120mm}}
\caption{Clustering as a function of centrists' threshold $dl$.
Vertical bars represent clusters: their center gives cluster's
average opinion and their length give the number of 
agents belonging to the cluster. Note the existence
of e.g. centrist attractors at $dl=0.28$, 
 single extreme attractors at $dl=0.35$,
double  extreme attractors at $dl=0.56$.}
\end{figure}

  For low extremist density, we observed the same 3 regimes, centrism,
polarised  extremism and bi-extremism, as in the full mixing topology
(figure 5).

 At larger threshold, $dl>0.5$ bi-extremism is always observed
(see e.g. figure 6).
In the presence of several extremists at both ends of the spectrum, the
lattice is divided in extremist domains of different opinions with
boundaries separated by opinion differences larger than the 
tolerance threshold. The size of these domains is smaller
than the lattice size and the dilution of extremism 
in the sea of centrist as observed for single extreme dynamics
at large $dl$ values (see figure 3) does not occur.
The evolution of the domains towards centrist opinion is limited
  by their size, which varies itself as the inverse of the square root
of extremist densities.   

   As soon as the threshold is lower than 0.5,
centrists position become stable. Their importance increases
when threshold decreases. In the intermediate region,
$0.25<d_l<0.5$, the 3 regimes can be observed.
 Which regime is observed depends of the initial sampling of the 
homogeneous opinion distribution. But, of course, the probability of observing
a given regime depends upon the threshold.  

These results apparently contradict \cite{ambdef}
 who don't report
the observation of single extremism attractor: in fact this is because
their simulation were done at higher extremist densities than ours.

 \begin{figure}[ht]
\centerline{
\epsfxsize=80mm
\epsfig{file=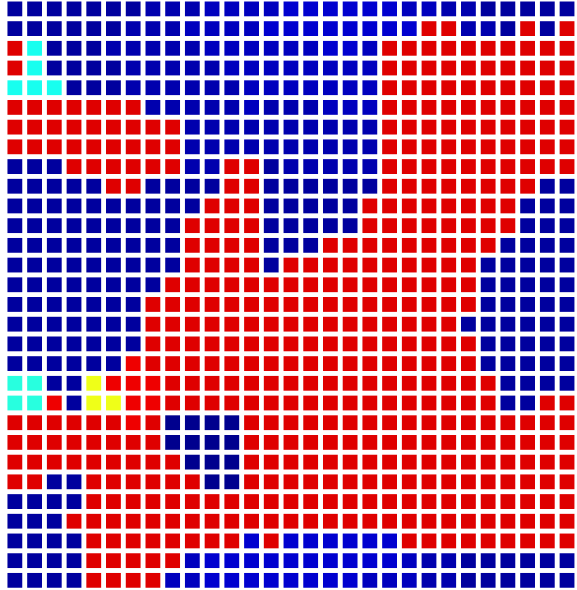,width=80mm}}
\caption{A bi-extremism opinion pattern observed after
100 iterations per agent for initial tolerances of $dl=0.65$
and $de=0.001$. Initially, twelve extremists (on each side 0 or 1)
were present. Two extreme domains are seen with 
a few agents at the interface with centrist opinion but low tolerance.}
\end{figure}

    Since  evolution towards a single extreme
is the most intriguing regime let us first observe
 the evolution towards this
regime. The next figures display opinions
and tolerances on a 64x64 square lattice after 300 and 2000 iterations
per site on average.  There were initially 12 extremists of either 
side, $d_l=0.38$, $d_e=0.01$ and $\mu=0.5$. 
Extremism, both in opinions and in intolerance,
 clearly propagates from the lower left side of the   
 lattice and eventually invades it.

\begin{figure}[ht]
\centerline{\def \epsfsize#1#2{0.8#1}\epsfbox{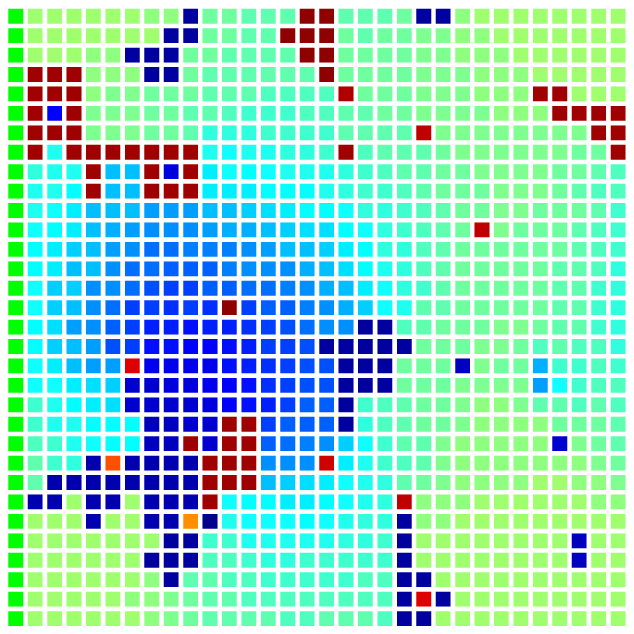}
\hskip 0.3 cm\epsfbox{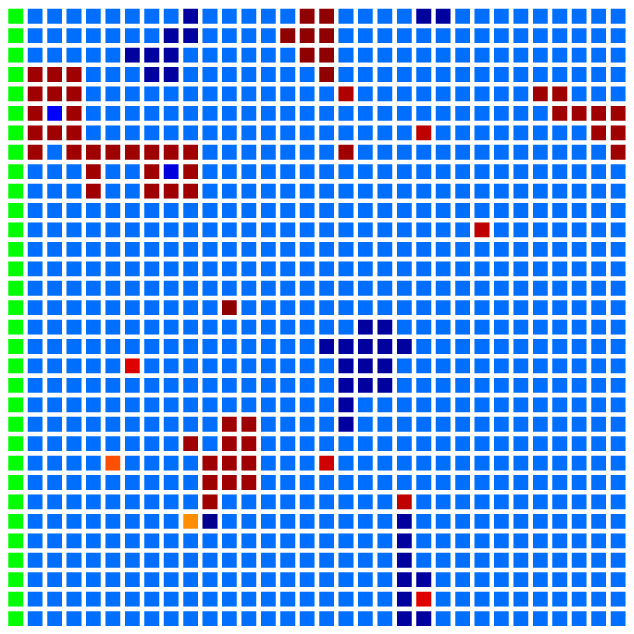}}
\vskip 1 cm
\centerline{\def \epsfsize#1#2{0.8#1}\epsfbox{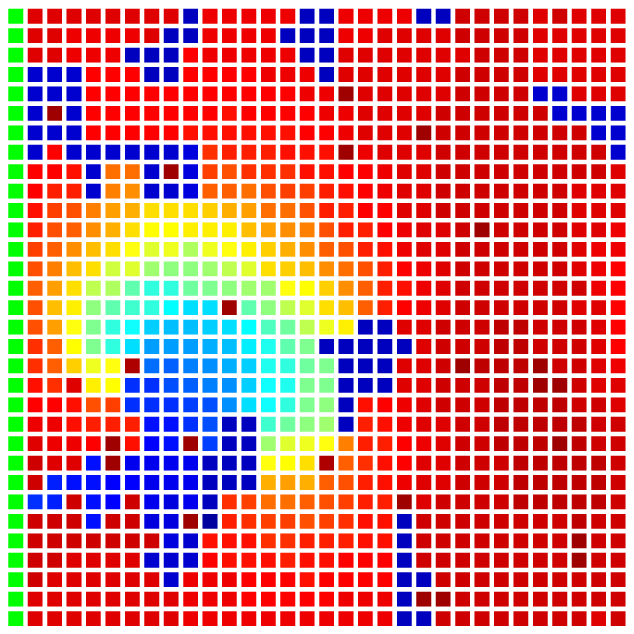}
\hskip 0.3 cm\epsfbox{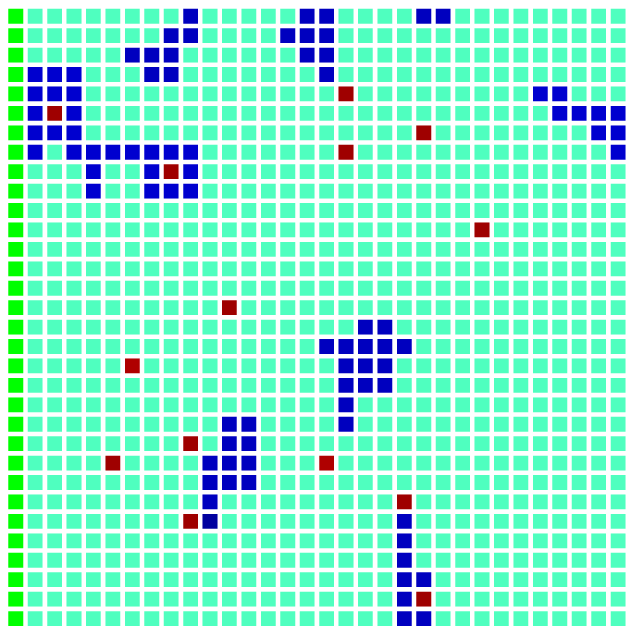}}
\caption{Upper frames: Opinions after 300 iterations per site 
(left panel) and 2000 (right panel). Color scale: deep blue 0,
light green 0.5, brown 1.
Lower frames: Tolerances after 300 iterations per site 
(left panel) and 2000 (right panel). Color scale: deep blue 0,
brown $d_l=0.38$, the maximum initial centrist's tolerance.}
\end{figure}

 After 2000 iterations the percolating cluster is uniform
at a low, but not extreme, opinion and tolerance values.

The patterns of figure 7 exemplify the different possible influence
of initial extrmists, according to the first events occuring
in their immediate neighborhood. One can distinguish
two different ``geographic'' configurations:
\begin{itemize}
\item Mesas
A few  extremist islands, with opinion and tolerance
close to the initial extremist values
survive, but their influence on their neighborhood is zero:
the difference in opinion at the edge between those
agents on the mesa and all their neighbours is larger
than $dl=0.38$. Obviously small values of $dl$ favour
mesa which disappear whenever $dl>0.5$.
\item Hills
  The success of the extremists in the lower left
corner is due to  local
fluctuations in opinions: there exists
in their neighborhood centrists which opinion is
within the large tolerance distance. Since the dynamics
results in decreasing local opinion gradients, the
diffusion process once started carries on across the lattice,
unless such hills collide at larger initial extremist densities
as seen on figure 6. 
\end{itemize}

In fact restricting interactions to the lattice structure
results in two dynamics: fast local opinion clustering
 and eventually slow diffusion 
across the lattice of local fluctuations.
A time plot of opinion and gradients average  across the lattice 
allow to distinguish among the fast local averaging of opinions
and the slow diffusion of extremism (figure 8). The average 
squared opinion gradient is evaluated by :
\begin{equation}
  G=\sum_i{(x_{i+1}-x_i)^2}
\end{equation}

The average gradient decreases very rapidly towards low values 
reflecting the fact that opinions locally average very fast.
This fast relaxation time does not depend upon the size
of the lattice.

The slow diffusion of extremism is reflected in the slow decrease of 
the average opinion from 0.5 the average of the initial uniform
distribution towards 0.25 corresponding to the 
"relative" overcome of extremism. The diffusion time varies as 
the square of the linear size of opinion domains.

 \begin{figure}[ht]
\centerline{
\epsfxsize=80mm
\epsfig{file=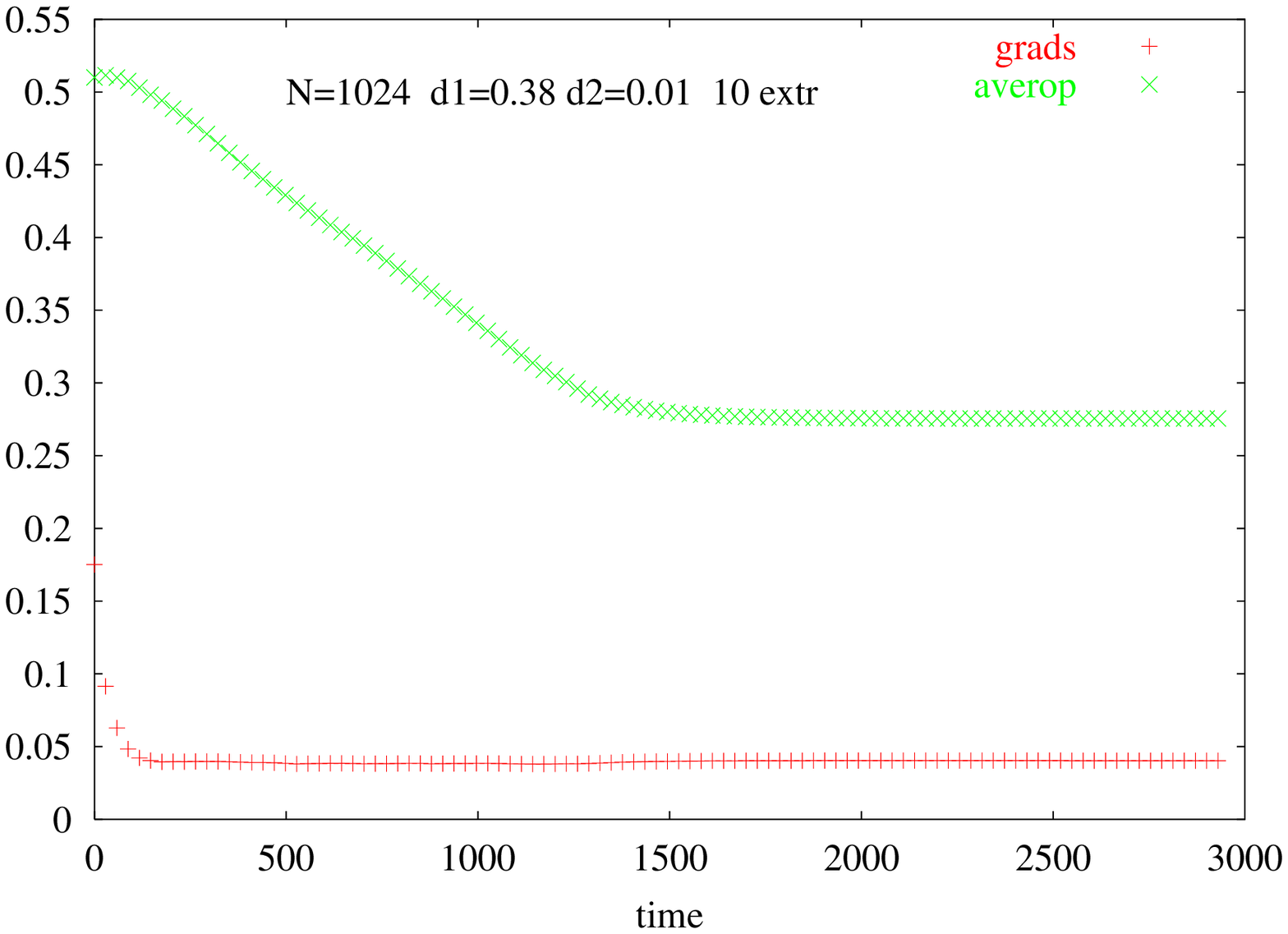,width=120mm}}
\caption{Time evolution of opinion and gradients average across the lattice.}
\end{figure}

\section{The global picture and the "hopeful monster hypothesis"}

  Our expectation which global attractor is reached when extremists of either side
are randomly scattered is then:

  \begin{itemize}
  \item Above $d_l=0.5$, extremists are always able to influence some
centrists in their immediate neighborhood and extremism
prevails at large times. When extremists of both kinds are present
 the lattice ends up subdivided in domains of extremists of either 
side.
\item Between $d_l=0.5$ and $d_l=0.25$, 
extremists are not always able to influence
centrists in their immediate neighborhood. Chances of conversion to extremism
depend upon the existence of neighbours close enough
in opinion (i.e. with difference in opinion smaller than threshold).
  Since an initial extremist is only one among four neighbours 
of its centrist neighbours, the random sampling of pairs might result 
in initial interactions of these neighbours with centrists 
(with a 0.75 probability), thus making it harder
 for the extremist and its neighbour to later interact.  
\begin{itemize}
\item At small extremist density, there is a threshold region such that 
the probability of having only one diffusing extremist "hill" is large enough
to observe single extremism convergence. This is also true
if there are several "hills" on the same extremist side, either close to 0 or
close to one.
\item At large extremist density, one obtains several hills
and bi-extremism is by far the most frequent attractor. 
\end{itemize}
Chances of extremists to influence their neighbours anyway decrease with 
 $d_l$ and one observes mostly one centrist attractor when   $d_l<0.33$.
\item Below $d_l=0.25$, the lattice is highly divided between many clusters
some of which are extremists.
  \end{itemize}

 The basic hypothesis is that when the density of extremists is low, 
 the initial growth of "extremist hills" are independent events
which occurrence only depends upon a restricted neighborhood of 
each initial extremist. These "extremist hills" can be called 
"hopeful monsters" since they are susceptible to grow
and invade the lattice as opposed to the "mesa" configurations.

 If we call $P_0$ the probability of occurrence of a
hopeful monster, we can obtain the probabilities of
observing any of the 3 attractors by simple combinatorics.
In the presence of $2n_e$ initial extremists 
($n_e$ extremists close to 1, $n_e$ close to 0) on a
large lattice (to ensure independence), 
these probabilities are given by:
\begin{eqnarray}
  P_c = Q_0^{2n_e}\\
  P_{bie} = (1-Q_0^{n_e})^2\\
  P_{moe} = 2 (1-Q_0^{n_e}) Q_0^{n_e}
\end{eqnarray}

 \begin{itemize}
   \item  where: $Q_0=1-P_0$ is the probability of any initial extremist
to give a "sterile" mesa; 
    \item $P_c$ is the probability of getting
a centrist cluster (i. e. du to the absence of any extremist hill);
\item $P_{bie}$ is the probability of getting
clusters of extremists of both kind (two kinds of hill present);
\item $P_{moe}$ is the probability of getting
a single extremist cluster (only hills of the
same extreme grow);
    \end{itemize}
  These expressions are immediately generalised 
to the asymmetric case when the initial numbers of 
extremists close to 0 and to 1 are different. 
They imply that the initial number of extremists
is important, not their density, at least in the limit
of low densities.

 The exact calculation of  $P_0$ as a function of the threshold $d_l$
involves a rather intricate combinatorics on the possible
initial configurations of 
the extremist's neighborhood and on the initial sequence of iterations
. But $P_0$ is easily evaluated by simulations. 
We did it on a 32x32 square lattice with a single extremist in the center.

\begin{figure}[ht]
\centerline{
\epsfxsize=80mm
\epsfig{file=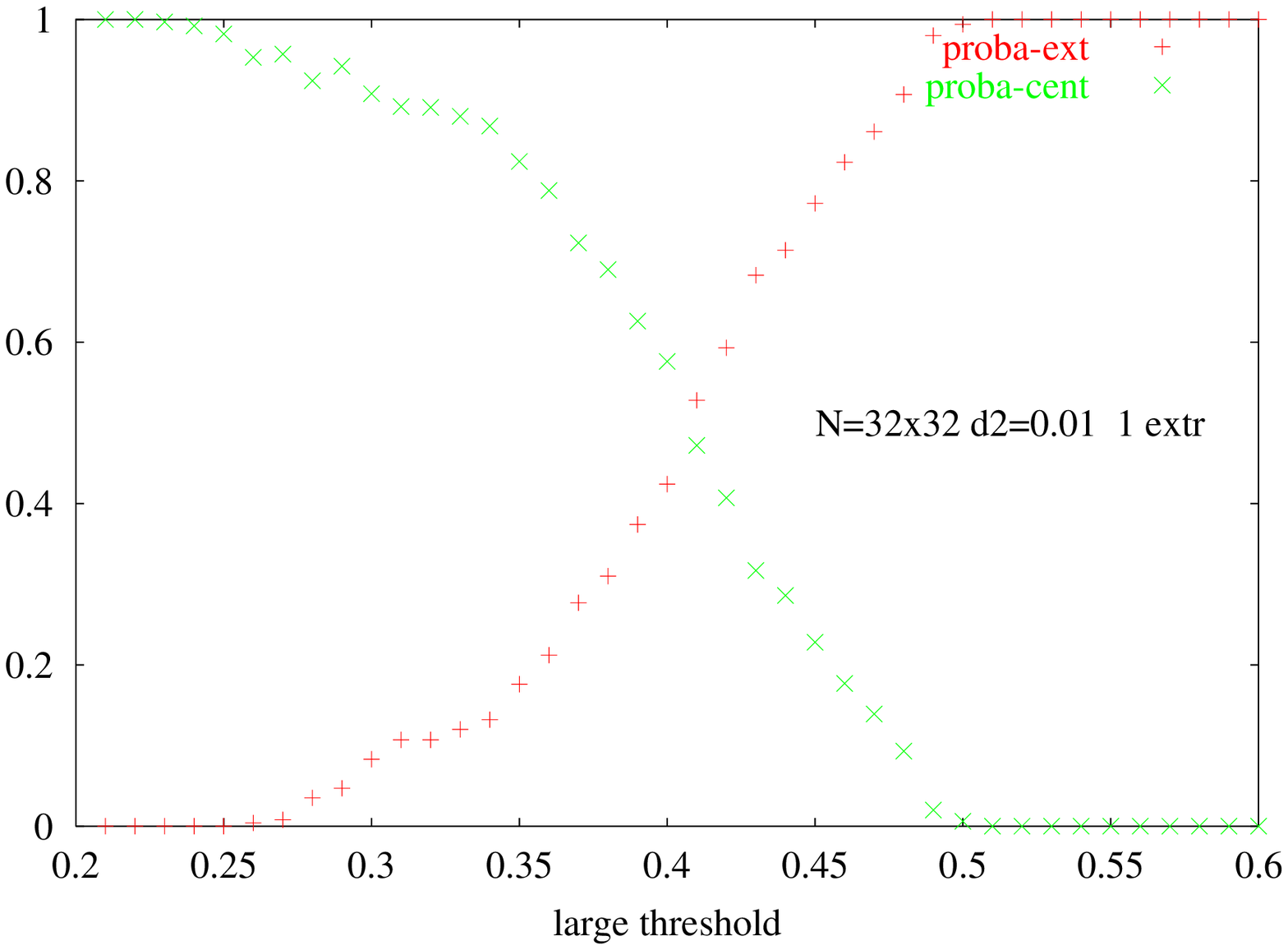,width=120mm}}
\caption{Statistics of  
extremist attractors (red '+')  and centrists attractor
(green x) as a function of centrists' threshold.
Each point corresponds to 1000 samples on
a 32x32 lattice with one central extremist.}
\end{figure}

Knowing  $P_0$  allows to check the "independent hopeful monster hypothesis"
which predicts the occurrence of attractors with probabilities given
by equations (8-10). Let 's take the case of  $n_e=3$.
Equation 5 predicts a maximum $P_{moe}$ probability of occurrence
of a single extremist attractor of 0.5 at $Q_{0}^3=0.5$, which corresponds 
to  $Q_{0}=0.79$ and approximately to $d_l=0.35$ according to figure 5.
 The statistics plotted on figure 6 roughly confirm this prediction:
The maximum of $P_{moe}$ is around  0.5 and occurs around $d_l=0.38$.

\begin{figure}[ht]
\centerline{
\epsfxsize=80mm
\epsfig{file=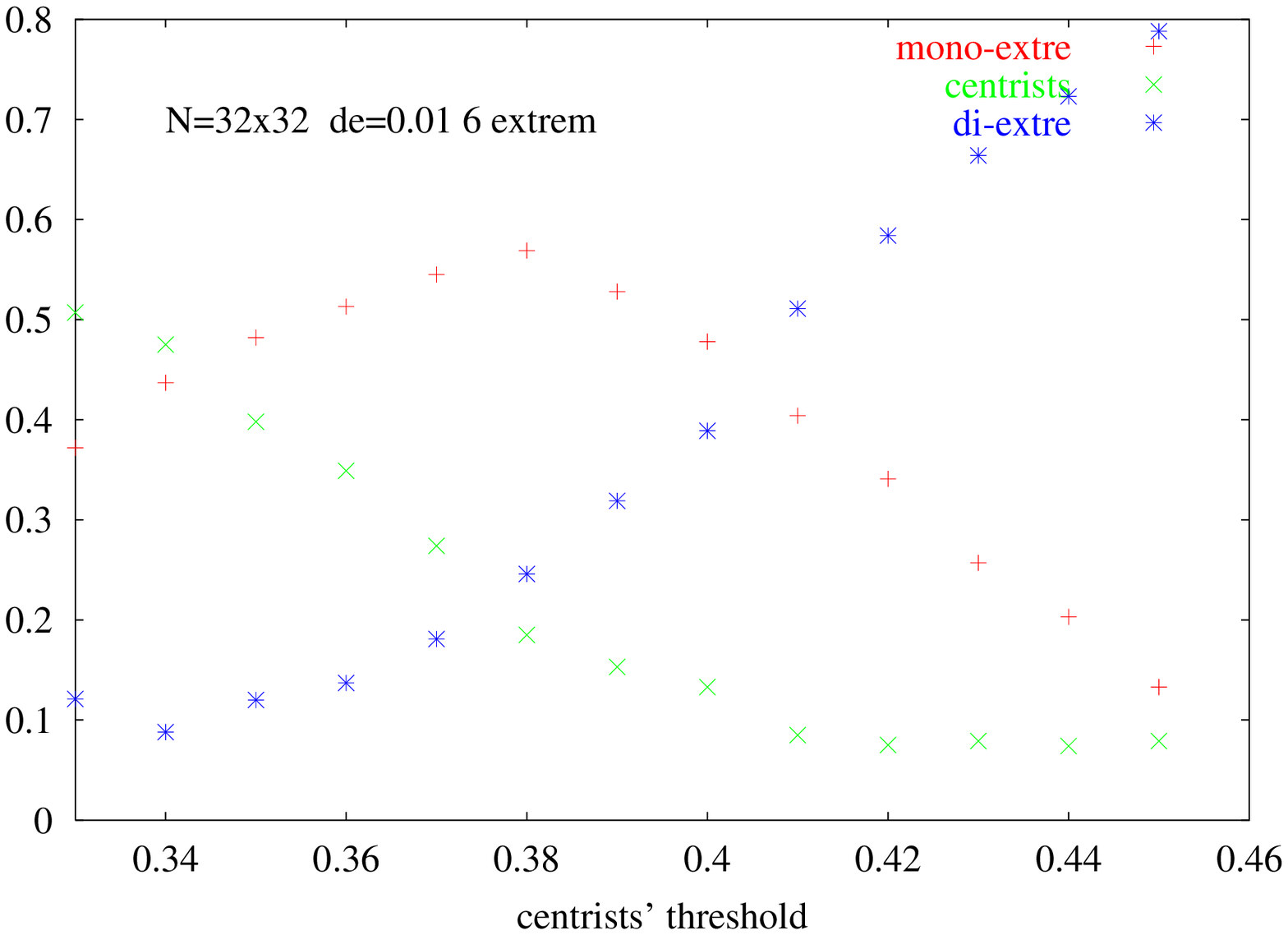,width=120mm}}
\caption{Statistics of  
attractors: red '+' are single extreme, 
blue '*' are double extreme and
green 'x'are centrist attractors; as a function of centrists' threshold.
Each point corresponds to 1000 samples on
a 32x32 lattice with 6 extremists.}
\end{figure}

  Figure 9 and expressions 8-10 then give a clear prediction
of the succession of the most frequent attractors when 
centrist initial tolerance is decreased from 0.5 to 0.25.
\begin{itemize}
\item 
Bi-extremism is predominant until $(1-P_0)^{n_e}$ is close 
to 0.5. How close?
\item 
 The width of the single extremist
region, $W_{moex}$, is evaluated from equation 10. We define it as the 
region where the probability of the single extremist
attractor is above one half of the maximum ($P_{moe}>0.25$).
\begin{equation}
  W_{moex} \propto (n_e )^{-1}
\end{equation}
\item
  Below the single extremist
region, centrist attractors are predominant. 
\end{itemize}

  So according to the above analysis,
single extreme attractors should be observable even
in the presence of many initial extremists. 

But large
fluctuations of the statistics of "hills" and "mesa"
are observed in figure 9 and 10 due to the vicinity
 of regime transitions (these two figures represent averages
over 1000 samples, and fluctuations are still noticeable).
 These fluctuations reduce the occurence of single
extreme attractors at larger $n_e$ values. Furthermore
increasing $n_e$ decreases the distance between sources 
of intolerance that cannot be considered as independent anymore:
 the probability of a "centrist" to be early 
influenced by an extremist is increased by having more than one 
extremist neighbour.

A rapid survey of the  $d_l$ region most favourable to single
extreme attractor, $0.34<d_l<0.40$, when the number of extremists
is increased from 0.4 to 2 perc. show that the probability
of observing  single extreme attractors decreases from 50 perc. to 4 perc.
 This is consistent with Amblard and Deffuant \cite{ambdef} who report 
the absence of any single extreme attractor for 
extremist densities higher than 2.5 perc. 

\subsection{Scale free networks}

 What about more realistic
topologies? Since the successive neighbourhood structure is preserved in
all networks topologies, except fully connected networks,
 we expect that the same intermediate scale features
which drive the dynamics, such as mesa or hills in opinions
or boundaries across domains are present for different topologies.
Can we expect equivalent phase diagrams, with possibly
more irregularities such as outliers and co-existence phases?

  We then run the extremist dynamics on scale free networks\cite{BA}
to test the above prediction (equivalent phase diagram).
(After small worlds networks were introduced by Watts and Strogatz
\cite{wat}, scale free networks became recently 
the strongest contenders as models of social networks).
  Scale free networks differ from lattices by
the inhomogenity of connectivity and by their
smaller diameter.

  We used a standard construction method to generate
  scale free networks, see e.g. Stauffer
and Meyer-Ortmanns \cite{HMO}:

 Starting from a fully connected network of
3 nodes, we add iteratively nodes (in general up to
900 nodes) and connect
them to previously created nodes in proportion
to their degree. We have chosen to draw two
symmetrical connections per new added node in order to achieve
the same average connection degree (4) as in the 30x30 square
lattice taken as reference. But obviously the obtained networks
are scale free as shown by Barabasi and Albert\cite{BA}.

  In fact \sfn \cite{BA} display a lot of heterogeneity in nodes connectivity.
In the context of opinion dynamics,
well connected nodes might be supposed more influential,
but not necessarily more easily influenced. At least this is
the hypothesis that we choose here.
 We have then assumed asymmetric updating:
a random node is first chosen, and then one of
its neighbours. But only the first node in the pair
might update his position according to equ.1, not both. As a result,
well connected nodes are influenced as often as
 others, but they influence others in proportion
to their connectivity. This particular choice
of updating is intermediate between what
Stauffer and Meyer-Ortmanns \cite{HMO} call directed and
undirected versions.

The cluster diagram obtained with 24 initial extremists
out of 900 agents (with the same parameters as for figures
5 and 6) is represented on figure 11.

 \begin{figure}[ht]
\centerline{
\epsfxsize=80mm
\epsfig{file=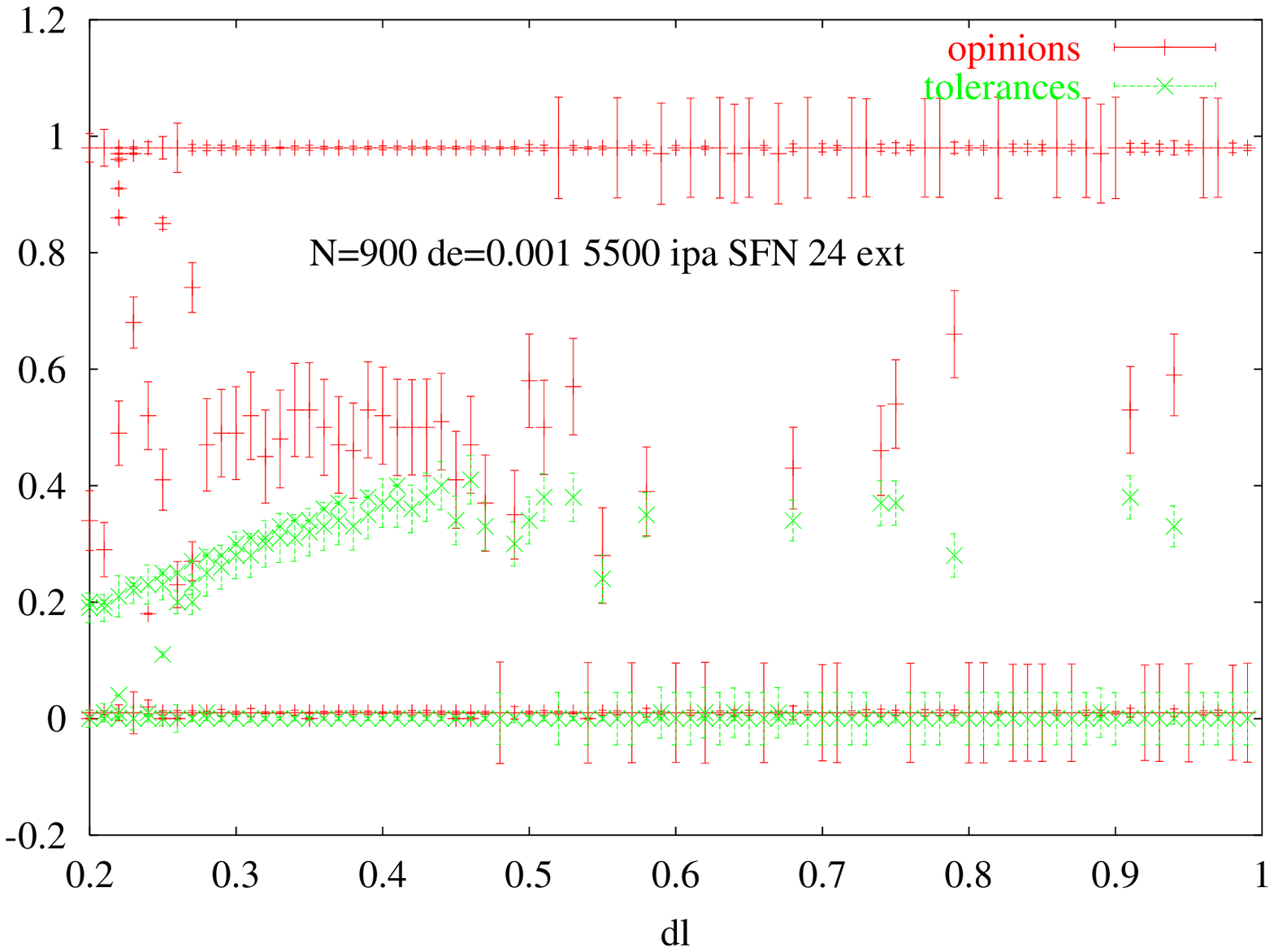,width=120mm}}
\caption{Clustering as a function of centrists' threshold $dl$
for a \sfn with the same parameters as for the diagram displayed 
for the square lattice on figure 5. The size of the bars representing
tolerance clusters is reduced by a factor two for clarity reasons.}
\end{figure}

\begin{itemize}
\item Similarity with lattice dynamics.
Below $dl=0.45$ this cluster diagram closely resembles 
those we obtained for square lattice, with predominance of 
centrism.
 Above $dl=0.45$ one also observes the predominance of
some kind of extremism, with a limited final tolerance.
\item Differences.
But the only two kinds of observed attractors are
low tolerance centrism and single sided extremism.
We don't observe two sided extremism attractors 
as with lattices.
The large inhomogeneity in nodes connectivity favour
well connected nodes: most often, the best connected extremist
impose his view nearly everywhere. And there are still
minority clusters around the other extremists.
  The asymptotic clusters with central opinion
are probably obtained when the initial sampling of 
extremists does not contain highly connected nodes.
\end{itemize}

The present result is still
preliminary: the distribution of the connectivity
of initial extremists is only a rough predictor
of the outcome of the dynamics. More complete studies,
outside the scope of the present paper are still needed.

\section{Conclusions}

 The above series of simulations give a clearer picture
of the phenomena occurring in this strongly simplified model of
opinion dynamics.
 
 The most important
result, already established in Deffuant etal \cite{dext}, is also
true for lattice and \sfn topologies:
 the existence of extremist regimes is largely due 
 to the large tolerance of agents which were
initially centrists.

  One difference is observed in the dilute regime on lattices:
when initial extremists are far apart in the network,
their influence at large tolerance $dl>0.5$
can be counter-balanced by centrists influence;
at infinite dilution centrism would win.
In some sense social networks can limit 
the propagation of extremism. Here is a 
possible explanation of the strategy    
of some sects which concentrate conversion efforts
on a limited number of individuals by repeated 
interaction rather than broadcasting across whole earth.

  We established that
all  attractors observed in the full mixing hypothesis,
 including single extreme, can be obtained on a square lattice
for low initial extremist densities.

 At higher extremist densities and large centrist tolerance,
the lattice structure favours two sided extremism,
while single sided extremism is favoured by 
scale free networks. Even our preliminary results allow
to understand why extremists (or market strategists)
should first convince
leader figures to establish their influence on a social network.

\par
 \vspace{1cm}
Acknowledgments:
   We acknowledge partial support from the FET-IST
grant of the EC IST 2001-33555 COSIN.
The present work was completed at the Institute for Scientific 
Interchange, Torino, which we thank for its hospitality.

\end{document}